\documentclass[conference]{IEEEtran}
\usepackage{mathrsfs}
\usepackage{url}
\usepackage{psfrag}
\usepackage{graphicx}
\usepackage{fancyhdr,epsfig}
\usepackage{epstopdf}
\usepackage{mathtools}
\usepackage{research5IEEE}
\usepackage{amsmath}
\usepackage{upgreek}
\usepackage{amsfonts}
\usepackage{amssymb}
\usepackage{amsthm}
\usepackage{tikz}

\newtheorem{theorem}{Theorem}

\newtheorem{remark}{Remark}

\allowdisplaybreaks[4]

\usetikzlibrary{decorations}

\newcommand{\e}{\textnormal{e}}

\newcommand{\fb}{\textnormal{fb}}

\newcounter{defcounter}
\title{A New Coding Scheme for Discrete Memoryless MACs with Common Rate-Limited Feedback}
\author{
\IEEEauthorblockN{Selma Belhadj Amor}
\IEEEauthorblockA{Telecom ParisTech, Paris, France\\
              {belhadjamor@telecom-paristech.fr}
              }                    
}
\begin{document}
 \maketitle  
\begin{abstract}
We propose a new coding scheme for the discrete memoryless two-user multi-access channel (MAC) with rate-limited feedback. Our scheme combines ideas from the Venkataramanan-Pradhan scheme for perfect feedback with ideas from the Shaviv-Steinberg scheme for rate-limited feedback.

Our achievable region includes the Shaviv-Steinberg achievable region and this inclusion can be strict.
For general MACs and for sufficiently large feedback rates, our scheme outperforms the Shaviv-Steinberg scheme as it achieves the same rate region as the Venkataramanan-Pradhan scheme for perfect feedback (which cannot be achieved by the Shaviv-Steinberg scheme). Furthermore, we numerically evaluate our achievable region with a specific (Gaussian) choice of random variables for the memoryless two-user Gaussian MAC. Our simulation results show that  
for some parameters of the Gaussian MAC and the feedback rate, our scheme achieves a strictly larger sum-rate than the Shaviv-Steinberg scheme.

%
\end{abstract}
\section{Introduction}
\label{sec:intro}
Gaarder \& Wolf~\cite{GaarderWolf} showed that perfect instantaneous output-feedback\footnote{By \emph{perfect instantaneous output-feedback} we refer to a model where each transmitter observes all previous channel outputs before it has to produce the next input. We henceforth refer to it also as \emph{perfect feedback.}} can increase the capacity of the two-user memoryless multiple-access channel (MAC) by enabling cooperation between the transmitters. The capacity region for general MACs with feedback is still unknown even for only two users. (A notable exception being Ozarow's capacity result for the two-user Gaussian MAC with perfect feedback \cite{Ozarow84}.) 

The Gaarder-Wolf scheme has been extended by Cover \& Leung \cite{CoverLeung} who introduced the ideas of block-Markov coding and superposition coding. Specifically, in the Cover-Leung scheme,  in each block~$b$, the transmitters send  independent fresh data superposed on common update information  belonging to the previous block~$(b-1)$. After observing the outputs in block~$b$, the receiver creates a list of all possible pairs of block-$b$ fresh data that is compatible (jointly typical) with these outputs. It also decodes the common update information. This common update information describes resolution information that allows the receiver to resolve its block-$(b-1)$ list, and thus to identify the fresh data that was sent in block~$(b-1)$. In order to be able to compute and send the block-$b$ common update information, the transmitters have to decode each other's fresh data sent in block~$(b-1)$ and calculate the receiver's block-$(b-1)$ list. They perform these tasks using their block-$(b-1)$ input 
signals and the block-$(b-1)$ feedback signals. (In case of perfect feedback, the latters correspond to the receiver's channel outputs.) For some MACs with perfect feedback the  presented Cover-Leung scheme is optimal and achieves capacity~\cite{Willems82}. For others, for example for the Gaussian MAC~\cite{Ozarow84}, it is strictly suboptimal~\cite{Ozarow84,BrossLapidoth, VenkatPradhan2011}. 

The Cover-Leung scheme has been improved by relaxing the requirement that after the transmission of each block, the transmitters have to decode each other's fresh data sent in this block~\cite{BrossLapidoth,VenkatPradhan2011}. Instead, the decoding at the transmitters (and also at the receivers) is delayed, allowing the transmitters to gain more information about each other's message before decoding. This results in less stringent rate-constraints as compared to the original Cover-Leung scheme. To implement this idea, Bross \& Lapidoth \cite{BrossLapidoth} proposed to append to each block a two-way transmitters-exchange phase and to delay the transmitters' decoding thereafter. 
Venkataramanan \& Pradhan \cite{VenkatPradhan2011} suggested to delay the transmitters decoding of the fresh data by an entire block. 
In their scheme, in each block~$b$ the transmitters send two sorts of resolution information, common \emph{receiver-side resolution information} to resolve the receiver's uncertainty about the block-$(b-1)$ fresh data, and correlated \emph{transmitters-side resolution information} to resolve each transmitter's uncertainty about the other transmitter's block-$(b-1)$ fresh data. 

Coding schemes were also presented for the MAC with generalized, noisy, or rate-limited\footnote{While  for generalized \cite{Carleial} or noisy feedback the feedback signals are ``passively" produced in a memoryless way from the channel inputs and outputs, in the model for rate-limited feedback the receiver can \emph{actively} code over the feedback links.} feedback. 
Carleial~\cite{Carleial} proposed a coding scheme for general discrete memoryless MACs with generalized feedback, which combines the Cover-Leung scheme with an optimal nofeedback scheme through rate-splitting. 
Lapidoth \& Wigger~\cite{LapidothWigger} proposed a scheme for the two-user Gaussian MAC with noisy feedback. Their scheme can be viewed as a robustification of Ozarow's capacity-achieving perfect-feedback scheme~\cite{Ozarow84} to noisy feedback. 

The main focus of this paper is on rate-limited feedback. For this model, 
Shaviv \& Steinberg~\cite{SteinbergShaviv} proposed a coding scheme based on Carleial's extension of the Cover-Leung scheme and on Heegard-Berger source coding~\cite{HeegardBerger} to communicate over the feedback links. For sufficiently large feedback rates their scheme achieves Cover \& Leung's achievable region for perfect feedback~\cite{CoverLeung} (which in this case coincides with Carleial's achievable region). 

In this paper, we propose a coding scheme for the two-user discrete memoryless MAC with common rate-limited feedback. Our coding scheme is based on the Venkataramanan-Pradhan scheme and on Heegard-Berger source coding~ \cite{HeegardBerger} over the feedback links. Our new  region includes the  Shaviv-Steinberg achievable region and this inclusion can be strict.
For sufficiently large feedback rates, our achievable region coincides with the  Venkataramanan-Pradhan achievable region.
%
\section{Channel Model}
\label{sec:channel}
We consider the two-user discrete memoryless MAC with rate-limited feedback. 
The setup is characterized by the triple of finite alphabets $(\set{X}_1,\set{X}_2,\set{Y})$, the conditional probability distribution $P_{Y|X_1X_2}$, and a nonnegative feedback rate $R_{\fb}$. At each time $t\in \mathbb{N}$, if $x_{1,t}\in \set{X}_1$ and $x_{2,t}\in \set{X}_2$ denote the signals sent by Transmitters~1 and~2, the receiver observes the channel output $y_t\in\set{Y}$ with probability
$P_{Y|X_1X_2}(y_t|x_{1,t},x_{2,t})$. 

The goal of communication is that Transmitters~1 and~2 convey the independent messages $M_1$ and $M_2$ to the common receiver. The messages $M_1$ and $M_2$ are uniformly distributed over 
$\mathcal{M}_1=\{1,\dots, \lfloor 2^{n R_1}\rfloor\}$ and $\mathcal{M}_2=\{1,\dots, \lfloor 2^{n R_2}\rfloor\}$, where $R_1$ and $R_2$ are the rates of transmission and $n$ is the blocklength. 

We assume common rate-limited feedback from the receiver to both transmitters. Specifically, upon observing $Y_t$, the receiver can send a feedback signal $F_{t}\in\set{F}_{t}$ to both transmitters where $\set{F}_{t}$ denotes the finite alphabet of $F_{t}$. 
The feedback signals $F_1,\ldots, F_n$ are of the form
\begin{IEEEeqnarray}{rCl}
F_{t}&=&\psi_{t}^{(n)}\left(Y_1,\dots,Y_t\right) ,\quad i\in \{1,2\}, \quad t \in \{1,\dots,n\},\label{eq:feedback_encoding}
\end{IEEEeqnarray}
for some feedback-encoding functions 
$\psi_{t}^{(n)}\colon \set{Y}^t \to \set{F}_{t}$.
It is assumed that both transmitters receive the feedback signals perfectly whenever the former satisfy the rate constraint on the feedback links: 
\begin{IEEEeqnarray}{rCl}
 |\set{F}_{1}|\times \ldots \times |\set{F}_{n}|&\leq& 2^{n R_{\fb}}.  \label{eq:feedback_constraint}
\end{IEEEeqnarray}
(The present feedback rate constraint is rather weak. One could imagine a stronger constraint where each sample $F_t$ has to satisfy $H(F_t)\leq R_{\fb}$. It can be easily shown that the two definitions are equivalent in terms of achievable rates.)
Notice that here the alphabets $\set{F}_1,\ldots, \set{F}_n$ are design parameters of the coding scheme.

Transmitter~$i$'s channel input at time~$t$, $X_{i,t},$ for  $i\in\{1,2\}$, can depend on Message $M_i$ and the prior feedback signals $F_{1},\dots,F_{t-1}$:
\begin{IEEEeqnarray}{rCL}\label{eq:MACinputs}
X_{i,t}=\varphi_{i,t}^{(n)} (M_i,F_{1},\dots,F_{t-1} ),\quad t\in\{1,\dots,n\},\label{eq:encoding}
\end{IEEEeqnarray}
for some encoding functions of the form $\varphi_{i,t}^{(n)}:\mathcal{M}_i\times\set{F}_{1}\times \dots\times\set{F}_{t-1}\to \set{X}_i.$

The receiver bases its guess of its desired messages on the output sequence $Y_1,\ldots,Y_n$. That is, it produces 
\begin{equation*}
(\hat{M}_1,\hat{M}_2)=\phi^{(n)}(Y_1,\ldots, Y_n),
\end{equation*} 
 for a decoding function ${\phi^{(n)}\colon \set{Y}^n \to\mathcal{M}_1\times \mathcal{M}_2}$. There is an error in the communication whenever $(\hat{M}_1,\hat{M}_2)\neq (M_1,M_2)$. The average probability of error is thus
\begin{IEEEeqnarray}{rCL} 
 \label{eq:error-mac}
 P_{\e}^{(n)}\triangleq\textnormal{Pr}\big[ (\hat{M}_1,\hat{M}_2) \neq (M_1,M_2) \big].
\end{IEEEeqnarray} 

We say that a rate pair $(R_1,R_2)$ is achievable over the MAC with common rate-limited feedback if there exists a sequence of encoding and decoding functions $\{\{\varphi_{1,t}^{(n)}\}_{t=1}^n,\{\varphi_{2,t}^{(n)}\}_{t=1}^n,\phi^{(n)}\}_{n=1}^\infty$ as described above, a sequence of feedback alphabets $\{\set{F}_{t}\}_{t=1}^n$ satisfying \eqref{eq:feedback_constraint}, and feedback-encoding functions of the form~\eqref{eq:feedback_encoding} such that $P_{\e}^{(n)}$ tends to zero as the blocklength $n$ tends to infinity.

When $R_\fb=0$, the feedback signals have to be deterministic and the setup is equivalent to a setup without feedback.
When $R_\fb\geq\log_2 |\set{Y}|$, the setup is equivalent to perfect-feedback.

\section{Achievable Region}
\label{sec:ach_reg}
\begin{theorem}[Achievable Region]
\label{thm}
Let  $\set{W}$, $\set{U}_1$, $\set{U}_2$, $\set{V}_1$, $\set{V}_2$, $\set{Y}_{12}$, $\set{Y}_{1}$, and $\set{Y}_{2}$ be arbitrary finite sets. Also, let  $(W,V_1,V_2,U_1,U_2,X_1, X_2, Y,Y_{12},Y_1,Y_2)$ and $(\tilde{W},\tilde{V}_1,\tilde{V}_2,\tilde{U}_1,\tilde{U}_2,\tilde X_1, \tilde X_2, \tilde Y,\tilde{Y}_{12},\tilde Y_1,\tilde Y_2)$ be two correlated random tuples over the product alphabets $\set{W}\times\set{U}_1\times\set{U}_2\times\set{V}_1\times\set{V}_2 \times \set{X}_1\times \set{X}_2 \times \set{Y}\times \set{Y}_{12} \times \set{Y}_1 \times \set{Y}_2$ satisfying the following two conditions:
\begin{enumerate}
\item The joint distributions of the two tuples coincide: 
\begin{equation}\label{eq:cond31}
P_{WU_1U_2V_1V_2X_1X_2YY_{12}Y_1 Y_2} =P_{\tilde W \tilde U_1\tilde U_2\tilde V_1\tilde V_2\tilde X_1\tilde X_2\tilde Y\tilde Y_{12}\tilde Y_1 \tilde Y_2},
\end{equation} 
and each of them factors as 
\begin{IEEEeqnarray}{rCl}\label{eq:cond32}
\lefteqn{P_{WU_1U_2V_1V_2X_1X_2YY_{12}Y_1 Y_2}}\qquad  \nonumber \\ & =& P_W P_{V_1V_2}  P_{U_1|WV_1} P_{U_2| WV_2}   \nonumber\\&&\hspace{0.5cm}\cdot P_{X_1| W U_1 V_1} P_{X_2| W U_2 V_2} P_{Y|X_1 X_2}\nonumber\\ & &\hspace{1cm}\cdot P_{Y_{12}|Y W} P_{Y_2|W Y Y_{12}} P_{Y_1|W Y Y_{12}}  
\end{IEEEeqnarray}
where $P_{Y|X_{1}X_2}$ describes the channel law of our MAC. 
\item Defining $\tilde{S}\eqdef (\tilde W, \tilde{V}_1, \tilde{V}_2, \tilde{Y}_{12})$, 
the joint distribution over both tuples factors as 
\begin{IEEEeqnarray}{rCl}\label{eq:cond33}
\lefteqn{P_{ W U_1U_2V_1V_2X_1X_2Y Y_{12} Y_1 Y_2\tilde S \tilde U_1\tilde U_2\tilde X_1\tilde X_2\tilde Y \tilde Y_1 \tilde Y_2}} \qquad\qquad \nonumber \\ & =& P_{\tilde S \tilde U_1 \tilde   U_2 \tilde X_1 \tilde X_2 \tilde Y \tilde Y_1 \tilde Y_2}  P_{V_1|\tilde S \tilde U_1}P_{V_2|\tilde S \tilde U_2}\nonumber \\ & &\hspace{1cm} \cdot P_{W U_1 U_2 X_1 X_2 Y_{12} Y_1  Y_2 Y|V_1 V_2}.\IEEEeqnarraynumspace
\end{IEEEeqnarray}
\end{enumerate}

All nonnegative rate pairs $(R_1,R_2)$ satisfying Constraints~\eqref{eq:ach_region} on top of next page are achievable.
\begin{figure*}
\begin{subequations}
\label{eq:ach_region}
\begin{IEEEeqnarray}{lCl}
 R_{\fb}&\geq&\max\bigr\{I(Y_{12};Y |W X_1),I(Y_{12};Y |W X_2)\bigr\}+ I(Y;Y_1|Y_{12} X_1 W)+ I(Y;Y_2|Y_{12} X_2 W)\label{eq:fb_constraint}\\
R_1&\leq &I(X_1;Y|SU_1U_2 X_2)\nonumber\\&&+\min \{I(U_1 ;Y_2 Y_{12}|\tilde S\tilde Y_2 \tilde U_2\tilde X_2 W V_2 U_2 X_2), I(W;Y|\tilde W \tilde Y)+I(U_1;Y| W V_1 V_2 U_2)+ I(V_1;Y|\tilde Y\tilde S \tilde Y_1 \tilde Y_2\tilde U_2 W V_2)\} \\
R_2&\leq &I(X_2;Y|SU_1U_2 X_1)\nonumber\\&&+\min\{I(U_2;Y_1Y_{12}|\tilde S \tilde Y_1\tilde U_1\tilde X_1 W V_1 U_1 X_1),I(W;Y|\tilde W \tilde Y)+ I(U_2;Y| W V_1 V_2 U_1)+I(V_2;Y|\tilde Y \tilde S \tilde Y_1 \tilde Y_2 \tilde U_1 W V_1)\}\\
 R_1+R_2&\leq &I(X_1X_2;Y|SU_1U_2)+I(W;Y|\tilde W \tilde Y)+ I(U_1 U_2;Y| W V_1 V_2)+I(V_1 V_2 ;Y|\tilde Y\tilde S\tilde Y_1 \tilde Y_2  W)\\
R_1+R_2&\leq &I(X_1X_2;Y|SU_1U_2)\nonumber\\&&+\min\{I(U_2;Y_1Y_{12}|\tilde S \tilde Y_1\tilde U_1\tilde X_1 W V_1 U_1 X_1),I(W;Y|\tilde W \tilde Y)+ I(U_2;Y| W V_1 V_2 U_1)+I(V_2;Y|\tilde Y \tilde S \tilde Y_1 \tilde Y_2\tilde U_1 W V_1)\}\nonumber\\&&+\min\{I(U_1 ;Y_2Y_{12}|\tilde S \tilde Y_2\tilde U_2\tilde X_2 W V_2 U_2 X_2),I(W;Y|\tilde W \tilde Y)+I(U_1;Y| W V_1 V_2 U_2)+ I(V_1;Y|\tilde Y\tilde S \tilde Y_1 \tilde Y_2\tilde U_2 W V_2)\}\IEEEeqnarraynumspace
\vspace*{-1mm}
\end{IEEEeqnarray}
\end{subequations}
\hrule
\vspace*{-5mm}
\end{figure*}
\end{theorem}
\begin{remark}Using time-sharing, it can  be shown that also the convex hull of the region described in Theorem~\ref{thm} is achievable. 
\end{remark}

\begin{remark}When choosing  $V_1=V_2=Y_{12}=Y_1=Y_2=W=\emptyset$, $U_1=X_1$, and $U_2=X_2$, the achievable region in Theorem~\ref{thm} reduces to the nofeedback capacity region.

When $R_{fb}\geq \log_2|\set{Y}|$, Constraint~\eqref{eq:fb_constraint} is always satisfied. In this case, we can choose $Y=Y_{12}$ and $Y_1=Y_2=\emptyset$ to mimic the setup with perfect feedback. 
If now we specialize the achievable region in Theorem~\ref{thm} to the choices  $U_1=X_1$ and  $U_2=X_2$, we recover the  achievable region of Venkataramanan \& Pradhan for perfect feedback~\cite{VenkatPradhan2011}.

Finally, when choosing  $V_1=V_2=\emptyset$, the tilded and the non-tilded tuples become independent and the achievable region in Theorem~\ref{thm} specializes to the set of all rate pairs $(R_1, R_2)$ satisfying
\begin{subequations}
\label{eq:v1v2empty}
\begin{IEEEeqnarray}{cCl}
 R_{\fb}&\geq&\max_{i\in\{1,2\}}\bigr\{I(Y_{12};Y |W X_i)\bigr\}\nonumber\\
 &&+ I(Y;Y_1|Y_{12} X_1 W)+ I(Y;Y_2|Y_{12} X_2 W)\\
R_1&\leq &I(X_1;Y|W U_1 X_2)+\min \{I(U_1;Y| W  U_2)\nonumber\\&&+I(W;Y),I(U_1 ;Y_2 Y_{12}|W X_2)\} \\
R_2&\leq &I(X_2;Y|W U_2 X_1)+\min\{I(U_2;Y| W  U_1)\nonumber\\&&+I(W;Y),I(U_2;Y_1Y_{12}| W X_1)\}\\
R_1+R_2&\leq &I(X_1X_2;Y|W U_1U_2)+\min\{I(U_2;Y| W  U_1)\nonumber\\&&+ I(W;Y),I(U_2;Y_1Y_{12}| W  X_1)\}\nonumber\\&&
+\min\{I(U_1;Y| W  U_2)+I(W;Y),\nonumber\\&&I(U_1 ;Y_2Y_{12}| W X_2)\}\IEEEeqnarraynumspace\\
R_1+R_2&\leq &I(X_1X_2;Y)
\end{IEEEeqnarray}
\end{subequations}
where now $P_{WU_1U_2X_1X_2Y Y_{12}Y_1 Y_2}$ factors as
\begin{IEEEeqnarray}{rCl}
\lefteqn{P_{WU_1U_2X_1X_2Y Y_{12}Y_1 Y_2}} \nonumber\qquad \\ & =&  P_W  P_{U_1|W} P_{U_2| W} P_{X_1|U_1 W} P_{X_2|U_2 W} P_{Y|X_1 X_2}\nonumber\\&&\hspace{1cm}\cdot P_{Y_{12}|Y W} P_{Y_1|Y W Y_{12}} P_{Y_2|Y W Y_{12}}   .  \IEEEeqnarraynumspace
\end{IEEEeqnarray}
This region contains the achievable region by Shaviv \& Steinberg~\cite{SteinbergShaviv}.
\end{remark}

\section{Outline of Coding Scheme}
\label{sec:coding}
Let $W, U_1, U_2, V_1, V_2, X_1, X_2, Y, Y_{12}, Y_1, Y_2, \tilde W, \tilde U_1, \tilde  U_2, \tilde V_1,\break \tilde V_2, \tilde X_1, \tilde X_2,\tilde Y, \tilde Y_{12}, \tilde Y_1,\tilde Y_2$ be as defined in  Theorem~\ref{thm} so that they satisfy~\eqref{eq:cond31}--\eqref{eq:cond33}.
Fix a  nonnegative rate pair $(R_1, R_2)$ that satisfies rate constraints~\eqref{eq:ach_region} with strict inequalities. Using for example Fourier-Motzkin Elimination, it can be shown that there must exist rates $R_1'>0, R_2'>0$ so that they satisfy $R_1> R_1'$ and $R_2> R_2'$ and the following nine conditions
\begin{subequations}
\begin{IEEEeqnarray}{cCl}
R_{\fb}&>&\max\bigr\{I(Y_{12};Y |W X_1),I(Y_{12};Y |W X_2)\bigr\}\nonumber\\&&+\sum_{i\in\{1,2\}} I(Y;Y_i|Y_{12} X_i W),\label{eq:rfb} \IEEEeqnarraynumspace\\
R_1-R'_1 &< &I(X_1;Y|SU_1U_2 X_2)\label{eq:r1-r'1} \\
R_2-R'_2 &< &I(X_2;Y|SU_1U_2 X_1)\label{eq:r2-r'2}\\
R_1\hspace*{-0.5mm}-\hspace*{-0.5mm}R'_1\hspace*{-0.5mm}+\hspace*{-0.5mm}R_2\hspace*{-0.5mm}-\hspace*{-0.5mm}R'_2\hspace*{-0.5mm} &< &I(X_1X_2;Y|SU_1U_2)\label{eq:r-r'}\\
R'_1&< & I(U_1 ;Y_2Y_{12}|\tilde S \tilde Y_2\tilde U_2\tilde X_2 W V_2 U_2 X_2)\label{eq:r'11}\IEEEeqnarraynumspace \\
R'_2&< & I(U_2;Y_1Y_{12}|\tilde S \tilde Y_1\tilde U_1\tilde X_1 W V_1 U_1 X_1)\label{eq:r'21}\IEEEeqnarraynumspace\\
		R'_1&< & I(W;Y|\tilde W \tilde Y)+I(U_1;Y| W V_1 V_2 U_2)\nonumber\\&&
		+ I(V_1;Y|\tilde Y\tilde S \tilde Y_1 \tilde Y_2\tilde U_2 W V_2)\label{eq:r'12}\\
		R'_2&< & I(W;Y|\tilde W \tilde Y)+ I(U_2;Y| W V_1 V_2 U_1)\nonumber\\&&+I(V_2;Y|\tilde Y \tilde S \tilde Y_1 \tilde Y_2 \tilde U_1 W V_1)\label{eq:r'22}\\
		R'_1 +R'_2&< & I(W;Y|\tilde W \tilde Y)+ I(U_1 U_2;Y| W V_1 V_2)\nonumber\\&&+I(V_1 V_2 ;Y|\tilde Y\tilde S \tilde Y_1 \tilde Y_2 W).\label{eq:r'1+r'2}
	\end{IEEEeqnarray}
\end{subequations}
Fix also a constant $R_0$ satisfying: 
\begin{IEEEeqnarray}{rCl}
R_0 &<& I(W;Y|\tilde{W},\tilde{Y}).\label{eq:r0}
\end{IEEEeqnarray}

We briefly describe a random code construction for which the average probability of error (averaged over codebooks, messages, and channel realizations) can be shown to tend to 0. A deterministic coding scheme achieving the same rates can then be obtained via standard arguments.

Our coding scheme is based on block-Markov and superposition coding, rate-splitting, sliding-window decoding, and Heegard-Berger coding on the feedback links. It extends over $B+2$ blocks. 
For each $i\in\{1,2\}$, split Message $M_i$ into $B$ submessages $(m_{i}^{(1)}, \ldots, m_{i}^{(B)})$ each of rate $R_i/B$. For $b\in \{1,\dots,B\}$, split each of these submessages into a pair  $(j_{i}^{(b)}, k_{i}^{(b)})$ of rates $R_i'/B$ and $(R_i-R_i')/B$, respectively. For blocks $b\in\{B+1,B+2\}$, set $j_1^{(b)}=k_1^{(b)}=j_2^{(b)}=k_2^{(b)}=1$.

For each block $b\in\{1,\dots,B+2\}$, at Transmitter $i$, for $i\in \{1,2\}$, the $j$-message $(j_i^{(b)})$ is transmitted using a feedback scheme and is going to be decoded at the other transmitter and the $k$-message $(k_i^{(b)})$ is transmitted without using the feedback and is decoded only at the receiver.

For each block $b\in \{1,\dots, B+2\}$, Transmitter~$i$, for $i\in\{1,2\},$ sends $\vect{x}_i^{(b)}(k_i^{(b)}|\vect{u}_i^{(b)}(j_i^{(b)}|\vect{v}_i^{(b)},\vect{w}^{(b)}))$.
The choice of the sequences $\vect{u}_i^{(b)}$, $\vect{v}_i^{(b)}$, and $\vect{w}^{(b)}$ is explained next.

After each block $b$, the receiver compresses the channel outputs $\vect{y}^{(b)}$ it observed for this block into $(\vect{y}_{12}^{(b)},\vect{y}_{1}^{(b)},\vect{y}_{2}^{(b)})$ using Heegard-Berger coding~\cite{HeegardBerger}. The receiver
uses the feedback links only once to send the same message $M_\fb^{(b)}$ which---as shortly explained ahead at the end of this section---is the triple $(r_{12}^{(b)},r_{1}^{(b)},r_{2}^{(b)})$, indices of bins containing the quantized output sequences $\vect{y}_{12}^{(b)}$, $\vect{y}_{1}^{(b)}$, and $\vect{y}_{2}^{(b)}$. Upon receiving $M_\fb^{(b)}$, Transmitter~$i$, for $i\in \{1,2\}$, reconstructs the sequence $\vect{y}_{12}^{(b)}$ by looking for a codeword $\vect{y}_{12}^{(b)}$ in bin $r_{12}^{(b)}$ jointly typical with $(\vect{w}^{(b)},\vect{x}_i^{(b)})$. Then, it looks for a codeword $\vect{y}_{i}^{(b)}$ in bin $r_{i}^{(b)}$ jointly typical with $(\vect{w}^{(b)},\vect{x}_i^{(b)},\vect{y}_{12}^{(b)})$.

For a given block~$b$, the messages $j_1^{(b)}$ and $j_{2}^{(b)}$ are at first transmitted using the $\vect{u}_1^{(b)}$- and  $\vect{u}_2^{(b)}$-codewords of this block $b$. In contrast to the schemes by Cover \& Leung \cite{CoverLeung} or by Carleial \cite{Carleial}, the transmitters do not immediately decode each-other submessages $j_{1}^{(b)}$ or $j_{2}^{(b)}$ after learning the $(\vect{y}_{12}^{(b)},\vect{y}_{1}^{(b)},\vect{y}_{2}^{(b)})$ signals of block $b$. Instead, they wait for another block, where they exchange information helping them in the decoding. Specifically, at the end of each block $b\in \{1,\dots, B+1\}$, each Transmitter~$i$ computes \emph{transmitter-side} resolution information $\vect{v}_i^{(b+1)}$ as a (randomized) function of its block-$b$ codeword $\vect{u}_i^{(b)}$ and some common information $\vect{s}^{(b)}$ which is known to both transmitters. (The common information $\vect{s}^{(b)}$ consists of the $\vect{w}^{(b)}, \vect{v}_1^{(b)}, \vect{v}_2^{(b)}, \vect{y}_{12}^{(b)}
$ sequences; the former three are explained in the sequel.) 
The resolution information $\vect{v}_i^{(b+1)}$ is then sent in block $b+1$. The initial sequences  $(\vect{v}_1^{(1)},\vect{v}_2^{(1)})$ are drawn i.i.d. according to $P_{V_1V_2}$ and are known to everyone. In each block $b\in \{1,\dots,B+2\}$, the sequences $(\vect{v}_1^{(b)},\vect{v}_2^{(b)})$  
are correlated, which makes that sending them can be more efficient than sending independent data. (In particular, Condition \eqref{eq:cond33} ensures that they have i.i.d. joint distribution $P_{V_1 V_2}$.)
 
After reception of $\vect{y}^{(b+1)}$, the receiver creates a list of the most likely $(j_1^{(b)}, j_2^{(b)})$ message pairs based on $\vect{y}^{(b)}$ and $\vect{y}^{(b+1)}$ (and all the information that it has for blocks $b$ and $b+1$). 

Upon observing the feedback outputs in block~${b+1}$, Transmitter~$i$, for $\in\{1,2\}$,  uses the sequences $\vect{y}_{12}^{(b)}$, $\vect{y}_i^{(b)}$, $\vect{y}_{12}^{(b+1)}$, and $\vect{y}_i^{(b+1)})$ (and all the information that it has for blocks $b$ and $b+1$) to estimate the decoder's list of highly-likely message-pairs $(j_1^{(b)}, j_2^{(b)})$. At the same time, it also estimates the other transmitter's $\vect{v}^{(b+1)}$-sequence and decodes the other transmitter's message $j^{(b)}$. (Notice that at this point, the receiver is hindered compared to the transmitters as it does know any of the $\vect{v}^{(b+1)}$-sequences.)
Conditions~\eqref{eq:r'11} and \eqref{eq:r'21} ensure that each transmitter decodes the other transmitter's $\vect{v}^{(b+1)}$-sequences  correctly with high probability.
Therefore, besides having an estimate of the decoder's list, Transmitter $i$ also has an estimate of the position of the correct message pair therein. Let the index $\lambda_i^{(b+2)}$ describe this position. If there is no such index or if it exceeds $\lfloor 2^{nR_0}\rfloor$, then the index $\lambda_i^{(b+2)}$ is chosen uniformly at random from the set $\{1,\dots,\lfloor 2^{nR_0}\rfloor\}$. 
If the feedback information and the $j$-messages were decoded correctly, with high probability we have $\lambda_1^{(b+2)}=\lambda_2^{(b+2)}$. We abuse notation and call this index $\lambda^{(b+2)}$. The two transmitters send this index jointly in block $b+2$ using a cooperation sequence $\vect{w}^{(b+2)}$ that plays the role of \emph{receiver-side} resolution information. For blocks $b\in\{1,2\}$, it is fixed and known to everyone.  

Upon receiving $\vect{y}^{(b+2)}$, the receiver decodes the $\vect{w}^{(b+2)}$ codeword and the index $\lambda^{(b+2)}$ based on $\vect{w}^{(b+1)}$,   $\vect{y}^{(b+1)}$, and $\vect{y}^{(b+2)}$, and uses $\lambda^{(b+2)}$ to identify the correct message-pair $(j_1^{(b)}, j_2^{(b)})$ within its list.
Condition~\eqref{eq:r0} ensures that $\vect{w}^{(b+2)}$ can be correctly decoded at the receiver. The receiver's list for block $b$ can be resolved by  $\vect{w}^{(b+2)}$ with high probability if Conditions~\eqref{eq:r'12}-\eqref{eq:r'1+r'2} are satisfied.
Thereafter, the receiver also  decodes with high probability the messages $k_1^{(b)}$ and $k_2^{(b)}$, encoded in the $\vect{x}_1^{(b)}$- and  $\vect{x}_2^{(b)}$-codewords, based on $\vect{y}^{(b)}$ if Conditions~\eqref{eq:r1-r'1}-\eqref{eq:r-r'} are true. 

To compress $\vect{y}^{(b)}$ at the end of block $b\in\{1,\ldots, B+1\}$, the receiver  looks for a sequence $\vect{y}_{12}^{(b)}$ jointly typical with $\vect{y}^{(b)}$ and the decoded sequence $\vect{w}^{(b)}$. Then, for $i\in \{1,2\}$, it looks for a sequence $\vect{y}_{i}^{(b)}$ jointly typical  with $(\vect{ y}_{12}^{(b)},\vect{y}^{(b)},\vect{w}^{(b)})$. Let $r_{12}^{(b)}$, $r_{1}^{(b)}$, $r_{2}^{(b)}$ denote the indices of the bins containing $\vect{y}_{12}^{(b)}$, $\vect{y}_{1}^{(b)}$, and  $\vect{y}_{2}^{(b)}$, respectively. The Heegard-Berger coding~\cite{HeegardBerger} and Constraint~\eqref{eq:rfb} ensure that with high probability, Transmitter~$i$ can reconstruct $(\vect{y}_{12}^{(b)},\vect{y}_i^{(b)})$, for $i\in \{1,2\}$.

\section{Example: the Gaussian MAC}
\label{sec:gaussian}
\begin{figure*}
	\addtocounter{equation}{-1}
	\setcounter{equation}{24}
	\begin{small}
		\begin{subequations}		
			\label{eq:gaussian}
			\begin{IEEEeqnarray}{rCl}
				R_f&\geq&C\left(\frac{\sigma^2}{\sigma_{12}^2}+\frac{P}{\sigma_{12}^2}- \frac{P \theta}{\sigma_{12}^2}  \left(1 - \alpha - \beta +\frac{ \beta^2 \theta 
					\lambda^2}{(1 - (1 - \alpha - \beta) \theta)}\right)\right)\nonumber\\
				&&
				+\sum_{i=1}^2 C\left(\frac{\sigma_{12}^2}{\sigma_{i}^2}  \frac{\sigma^2 \left(1-  (1 - \alpha - \beta) \theta\right) + 
					P \left((1 - (1 - \alpha - \beta) \theta)^2 - \beta^2 \theta^2 \lambda^2\right)}{(\sigma^2 + \sigma^2_{12})\left(1 -  (1 - \alpha - \beta) \theta\right) + 
					P \left((1 - (1 - \alpha - \beta) \theta)^2 - \beta^2 \theta^2 \lambda^2\right)}\right)\\
				R_1 &\leq& C\left(\frac{P (1-\theta)}{\sigma^2}\right)+\min\biggr\{C\left(\frac{P \theta\alpha}{\sigma^2+ \frac{\sigma_{12}^2\sigma_2^2 }{\sigma_{12}^2+\sigma_{2}^2}+ P (1-\theta)}
				+\frac{ P \theta\beta  (1+\lambda)} {\frac{\sigma^2_{12} \sigma^2_2}{\sigma^2_{12} + \sigma^2_2} + \sigma^2  + P (1-\theta)+P  \alpha \theta }\right) ,C\left(\frac{\alpha P \theta}{2 P(1-\theta)+\sigma^2}\right)\nonumber\\&&+C\left(\frac{4 P\theta (1-\alpha-\beta)}{\sigma^2+2 P(1-\theta) +2 P \theta (\alpha+\beta (1+\lambda))}\right)+C\left(\frac{\beta P \theta  (2 P (1-\theta)+\sigma^2)(1+\lambda) }{(2\alpha P \theta +2 P (1-\theta) +\sigma^2)(\sigma^2+2 P (1-\theta)+\alpha P \theta ) }\right)\biggr\},\\
				R_2 &\leq& C\left(\frac{P (1-\theta)}{\sigma^2}\right)+\min\biggr\{C \left(\frac{P \theta\alpha}{\sigma^2+ \frac{\sigma_{12}^2\sigma_1^2 }{\sigma_{12}^2+\sigma_1^2}+ P (1-\theta)}
				+\frac{ P \theta\beta  (1+\lambda)} {\frac{\sigma^2_{12} \sigma^2_1}{\sigma^2_{12} + \sigma^2_1} + \sigma^2  + P (1-\theta)+P  \alpha \theta }\right),C\left(\frac{\alpha P \theta}{2 P(1-\theta)+\sigma^2}\right)\nonumber\\
				&&+C\left(\frac{4 P\theta (1-\alpha-\beta)}{\sigma^2+2 P(1-\theta) +2 P \theta (\alpha+\beta (1+\lambda))}\right)+C\left(\frac{\beta P \theta  (2 P (1-\theta)+\sigma^2)(1+\lambda) }{(2\alpha P \theta +2 P (1-\theta) +\sigma^2)(\sigma^2+2 P (1-\theta)+\alpha P \theta ) }\right)\biggr\},\\
				R_1+R_2&\leq & C\left(\frac{ 2 P (1-\theta)}{\sigma^2} \right)+C\left(\frac{2 P \theta (2-\alpha-\beta (1-\lambda))}{2 P (1-\theta )+\sigma^2}\right),\\
				R_1+ R_2&\leq & C\left(\frac{ 2 P (1-\theta)}{\sigma^2} \right)+
				\sum_{i=1}^2				
				\min\biggr\{C\left(\frac{P \theta\alpha}{\sigma^2+ \frac{\sigma_{12}^2\sigma_i^2 }{\sigma_{12}^2+\sigma_i^2}+ P (1-\theta)}+\frac{ P \theta\beta  (1+\lambda)} {\frac{\sigma^2_{12} \sigma^2_i}{\sigma^2_{12} + \sigma^2_i} + \sigma^2  + P (1-\theta)+P  \alpha \theta }\right),C\left(\frac{\alpha\theta P}{2 P(1-\theta)+\sigma^2}\right)\nonumber\\
				&&+C\left(\frac{4 P\theta (1-\alpha-\beta)}{\sigma^2+2 P(1-\theta) +2 P \theta (\alpha+\beta (1+\lambda))}\right)+C\left(\frac{\beta P \theta  (2 P (1-\theta)+\sigma^2)(1+\lambda) }{(2\alpha P \theta +2 P (1-\theta) +\sigma^2)(\sigma^2+2 P (1-\theta)+\alpha P \theta ) }\right)\biggr\}, 
			\end{IEEEeqnarray}
		\end{subequations}
	\end{small}
	\vspace*{-2mm}
		\setcounter{equation}{24}
	\hrule
	\vspace*{-3mm}
\end{figure*}
\setcounter{equation}{12}
Consider a memoryless Gaussian MAC with symmetric input-power constraint $P$. The channel output is $Y=X_1+X_2+Z$, where $Z$ is  zero-mean Gaussian with variance $\sigma^2>0$.
It can be shown that our coding scheme in Section~\ref{sec:coding} and Theorem~\ref{thm} in Section~\ref{sec:ach_reg} hold also for this Gaussian MAC.

Inspired by \cite{VenkatPradhan2011}, we propose the following choices. 
Let $\alpha, \beta>0$ such that $\alpha+\beta <1$, $\theta \in [0,1]$, and $\sigma_{12},\sigma_1,\sigma_2>0$. 
Also, let $\lambda \in[-1,1]$ so that 
\vspace*{-1mm}
\begin{equation}
\label{eq:cond_lambda}
\lambda \leq \frac{P \theta \alpha}{\sigma^2+\sigma_{12}^2+ P \alpha \theta + 2 P (1-\theta)}.
\vspace*{-1mm} 
\end{equation}
Now, let $W$, $A_1$, $A_2$, $I_{X_1}$,  $I_{X_2}$, $\tilde W$, $\tilde A_1$, $\tilde A_2$, $\tilde I_{X_1}$, and $\tilde I_{X_2}$ be independent zero-mean standard  Gaussians, and independent thereof and independent of each other, let $Z_{12},\tilde{Z}_{12}$ be zero-mean Gaussians of same variance $\sigma_{12}^2$, $Z_{1},\tilde{Z}_{1}$ be zero-mean Gaussians of same variance $\sigma_{1}^2$, and  $Z_2,\tilde{Z}_2$ be zero-mean Gaussians of same variance $\sigma_2^2$,
and let $(\tilde V_1, \tilde V_2)$ be a centered bivariate Gaussian of covariance matrix 
$\begin{pmatrix}1&\lambda\\\lambda&1\end{pmatrix}$.\\
Define for $i \in \{1,2\}$,
\vspace*{-1.5mm}
\begin{small}
\begin{IEEEeqnarray}{rCl}
 \tilde U_i&\eqdef& \sqrt{\alpha} \tilde A_i+\sqrt{\beta} \tilde V_i+\sqrt{1-\alpha-\beta} \tilde W,\\
 \tilde X_i&\eqdef & \sqrt{P (1-\theta)} \tilde I_{X_i}+ \sqrt{P \theta}\tilde U_i,\\
\tilde Y_i&\eqdef & \tilde Y+ \tilde Z_i,\\
\tilde Y_{12}&\eqdef & \tilde Y+ \tilde Z_{12}.
\end{IEEEeqnarray}
\end{small}
\vspace*{-1mm}
\\
Furthermore, define 
\begin{small}
\vspace*{-1mm}
\begin{subequations}
\begin{IEEEeqnarray}{rCl}
 V_1&\eqdef&\xi_1 \left(\tilde U_1-\sqrt{\beta } \tilde V_1-\sqrt{(1-\alpha -\beta)}\tilde W\right)/\sqrt{\alpha }+\xi_2 f(\tilde S)\IEEEeqnarraynumspace\\
V_2&\eqdef&-\xi_1 \left(\tilde U_2-\sqrt{\beta } \tilde V_2-\sqrt{(1-\alpha -\beta)}\tilde W\right)/\sqrt{\alpha }-\xi_2 f(\tilde S) \IEEEeqnarraynumspace
\vspace*{-1mm}
\end{IEEEeqnarray}
\end{subequations}
\end{small}
\vspace*{-2mm}
\\
where the function $f$ is chosen as 
\begin{small}
\begin{IEEEeqnarray}{rCl}
f(\tilde S)&=& f(\tilde W, \tilde V_1, \tilde V_2, \tilde Y_{12})\nonumber\\
&\eqdef&\frac{\tilde Y_{12}-\sqrt{\beta \theta P} (\tilde V_1+\tilde V_2)- 2 \sqrt{(1-\alpha-\beta)\theta  P} \tilde W}{\sqrt{\sigma^2+\sigma_{12}^2+2 P \alpha \theta + 2 P (1-\theta)}}\IEEEeqnarraynumspace
\end{IEEEeqnarray}
\end{small}
and where $\xi_1,\xi_2 \in \Reals$ are chosen to satisfy
\begin{subequations} 
\begin{small}
\begin{IEEEeqnarray}{rCl}
1&=& \xi_1^2+\xi_2^2+2\xi_1\xi_2 \sqrt{\frac{P \theta \alpha}{\sigma^2+\sigma_{12}^2+2 P \alpha \theta + 2 P (1-\theta)}},\IEEEeqnarraynumspace\label{eq:cond1}\\
\lambda&=&-2 \xi_1 \xi_2  \sqrt{\frac{P \theta \alpha}{\sigma^2+\sigma_{12}^2+2 P \alpha \theta + 2 P (1-\theta)}} -\xi_2^2.
\end{IEEEeqnarray}
\end{small}
\end{subequations}
\vspace*{-2mm}
\\(Condition~\eqref{eq:cond_lambda} ensures that such real $\xi_1$ and $\xi_2$ exist. In general, there  are four possible choices for $\xi_1,\xi_2$. The specific choice of $\xi_1, \xi_2$ does not show up in the rate-constraints~\eqref{eq:gaussian} and  does not change the set of achievable rates.)

For these choices define for $i \in \{1,2\}$,\vspace*{-1mm}
\begin{small}
\begin{IEEEeqnarray}{rCl}
 U_i&\eqdef& \sqrt{\alpha} A_i+\sqrt{\beta}  V_i+\sqrt{1-\alpha-\beta} W,\\
  X_i&\eqdef & \sqrt{P (1-\theta)}  I_{X_i}+ \sqrt{P \theta} U_i,\\
 Y_i&\eqdef & Y+ Z_i,\\
 Y_{12}&\eqdef & Y+ Z_{12}.
\vspace*{-1mm}
\end{IEEEeqnarray}
\end{small}
\indent Substituting the above choice into the rate-constraints of Theorem~\ref{thm}, we obtain that all nonnegative rate pairs $(R_1,R_2)$ satisfying Constraints~\eqref{eq:gaussian} on top of this page are achievable.
In~\eqref{eq:gaussian} we use the notation $C(x)\eqdef \frac12 \log(1+x)$.
\begin{figure}
\begin{center}
\includegraphics[height=5cm,width=8.5cm]{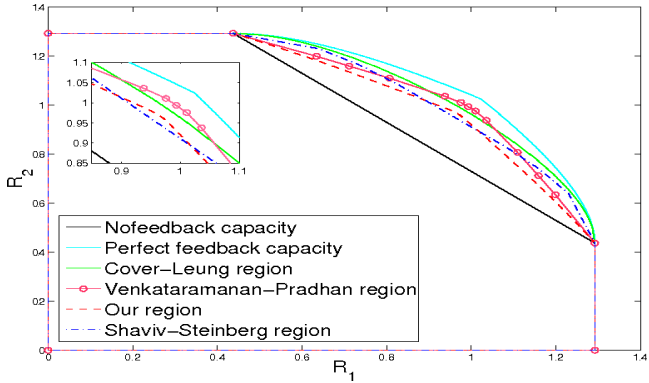}
 \end{center}
\vspace*{-5mm}
\caption{Achievable regions for the Gaussian MAC for $P/\sigma^2=5$ and $R_\fb=2$\label{figures2}}
\vspace*{-5mm}
 \end{figure}
 
Figure~\ref{figures2} compares the achievable region in~\eqref{eq:gaussian}, to the nofeedback capacity region, the perfect feedback capacity region~\cite{Ozarow84}, the Cover-Leung~\cite{CoverLeung} and Venkataramanan-Pradhan~\cite{VenkatPradhan2011} regions for perfect feedback, and to the Shaviv-Steinberg region with rate-limited feedback~\cite{SteinbergShaviv}.

For the sake of simplicity we restrict to the case where only common feedback is present ($Y_1=Y_2=\emptyset, \sigma_1^2,\sigma_2^2\to \infty$) which reduces to Wyner-Ziv coding~\cite{WynerZiv} over the feedback links. In this case, we need to have $\sigma_{12}^2 \geq \frac{\sigma^2+P}{2^{2R_\fb}-1}$ and we see that our scheme is strictly better in terms of sum-rate than the Shaviv-Steinberg scheme. 
In fact, based on extensive simulations, we conjecture that this is the case whenever 
  $\frac{P}{\sigma^2} < 2^{2 R_\fb}-2$,
which is equivalent to $\sigma_{12}^2 < \sigma^2$.
\section*{Acknowledgements}
The author would like to thank M.~Wigger for helpful discussions and the city of Paris for supporting this work under the ``Emergences'' program. 
\addcontentsline{toc}{chapter}{\numberline{}Bibliography}

\end{document}